\documentclass[11pt,twoside]{article}
\usepackage{asp2010ed}
\usepackage{subfigure}

\resetcounters

\begin{document}

\title{The Morpho-kinematics of Planetary Nebulae with Binary Central Stars}
\author{David Jones$^{1,2}$, Miguel Santander-Garc\'ia$^{3}$, Henri M.J. Boffin$^1$, Brent Miszalski$^{4,5}$ and Romano L.M. Corradi$^{6,7}$
\affil{$^1$European Southern Observatory, Alonso de C\'ordova 3107,  Santiago, Chile}
\affil{$^2$Universidad de Atacama, Copayapu 485, Copiap\'o, Chile}
\affil{$^3$Observatorio Astron\'omico Nacional, Ap 112, 28803 Alcal\'a de Henares, Spain}
\affil{$^4$South African Astronomical Observatory, P.O. Box 9, Observatory, 7935 Cape Town, South Africa}
\affil{$^5$Southern African Large Telescope Foundation, P.O. Box 9, Observatory, 7935 Cape Town, South Africa}
\affil{$^6$Instituto de Astrof\'isica de Canarias, E-38200 La Laguna, Tenerife, Spain}
\affil{$^7$Departamento de Astrof\'isica, Universidad de La Laguna, E-38206 La Laguna, Tenerife, Spain}
}

\begin{abstract}
It is now clear that a binary formation pathway is responsible for a significant fraction of planetary nebulae, and this increased sample of known binaries means that we are now in a position to begin to constrain their influence on the formation and evolution of their host nebulae.  Here, we will review current understanding of how binarity influences the resulting nebulae, based on observations and modelling of both the central binary systems and the planetary nebulae themselves.  We will also highlight the most important test-cases which have proved the most interesting in studying the evolution of binaries into and through the planetary nebula phase.
\end{abstract}

\section{Introduction}
Central star binarity has long been invoked as a source of angular momentum for the formation of aspherical and axisymmetric shapes in planetary nebulae (PNe).  But, until recently, too few binary central stars were known to be able to begin to draw strong conclusions on their exact influence on the final morphokinematical properties of the host nebulae.  Now, with a significant number of central binaries known and a selection with well-constrained binary properties (see Hillwig, these proceedings), it is now possible to begin to relate the properties of the nebulae to those of their central binaries, and reconstruct their evolution.

\section{The great leap forward}
\begin{figure}
\includegraphics[width=\textwidth]{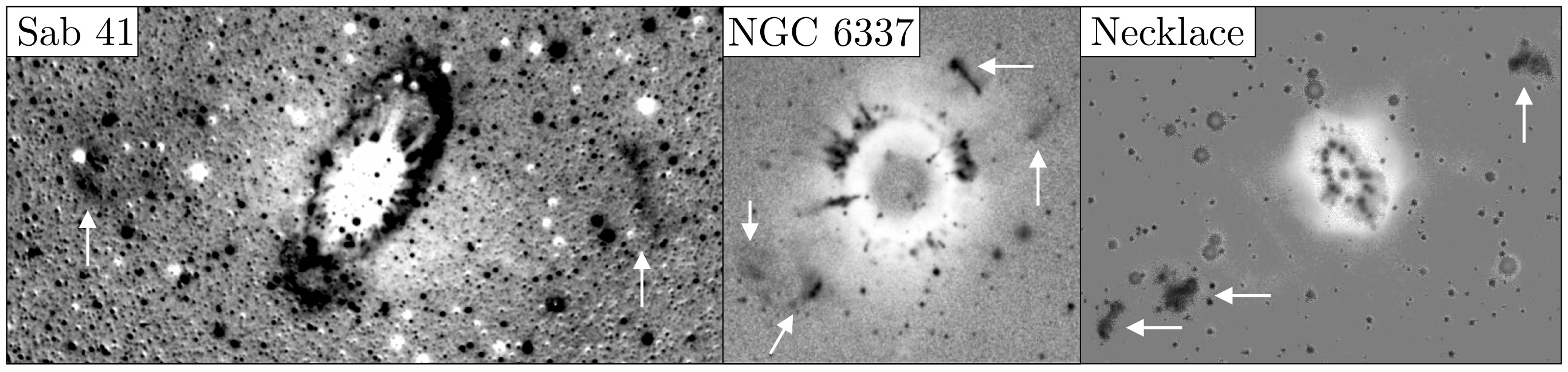}
\caption{Examples of PNe with binary central stars displaying the morphological traits (rings, jets and low-ionisation structures) found  by \citet{miszalski09b} to be prevalent amongst PNe with post-CE central stars.  Image reproduced from \citet{miszalski11c}.}
\label{fig:jetsski}
\end{figure}

Until as recently as 2009, as few as 16 post-common-envelope (post-CE) binary central stars were known, but with the advent of variability surveys (particularly the OGLE survey) this number increased dramatically \citep{miszalski09a}.  At this point, with $\sim$40 central binaries known, it was possible to begin compare the morphological properties of this sample to the wider PN population and identify traits which appeared more common in those nebulae hosting binaries (and may therefore be considered indicative of binarity).  \citet{miszalski09b} acquired new deep images of many of the sample and, combining them with archival imagery of a suitable standard (many previous investigations were based on, by today's standards, poor quality, low-resolution imagery), identified several morphologies that were more prevalent amongst their sample of binaries.  These morphologies included bipolarity, as would be expected, as well as polar-outflows/jets, low-ionisation filaments and equatorial rings (see figure \ref{fig:jetsski}).

This led the authors to begin an extensive programme to monitor, both photometrically and spectroscopically, the central stars of PNe displaying these traits in order to further augment the number of known binaries (and where possible characterise their properties).  This programme has borne fruit in the form of several new discoveries \citep{corradi11,miszalski11a,miszalski11b,boffin12,jones14}, including elusive eclipsing systems which are important for determining all the parameters of the central star system.

\section{Linking morphology to binarity}

Beyond just associating binary central stars with particular morphological traits, it has been possible to delve deeper into the problem and probe the evolution of these systems by combining our knowledge of the individual morphologies and parameters of the host binaries.

\subsection{Morphokinematical modelling}
One of the clearest predictions of the theories of PN shaping by a central binary are that the resulting nebula will have an elongated/bipolar morphology, so it is reassuring that the sample of known binary PN do indeed show a prevalence of bipolar morphologies.  However, this in itself is not enough to confirm that the binary was the shaping agent, rather it is important to examine a further prediction of these models by showing that the orbital plane of the binary coincides with the waist of the bipolar structure \citep[i.e. that that inclination of the orbital plane is perpendicular to that of the nebular symmetry axis][]{nordhaus06}.  This necessitates knowledge of both nebular structure and inclination as well as the orbital inclination of the binary, neither of which are trivial to discern.

Very few of the known binaries have well-constrained orbital inclinations as this requires detailed modelling and, in general, radial velocity measurements throughout the orbit.  However, a reasonable selection have been subjected to such modelling and are therefore useful in examining the relationship between orbital plane and nebular symmetry axis.  But, the determination of the inclination of the nebular symmetry axis is similarly non-trivial, as the orientation and morphology of a PN cannot be determined from imagery alone.  This arises because of a clear degeneracy between shape and inclination, where, for example, a bipolar nebula viewed ``end-on'' will appear as circular/spherical.  The only way to resolve this degeneracy is by combining imagery with spatially-resolved high-resolution spectroscopy of the nebula, and modelling the three-dimensional morpho-kinematic nebular structure \citep[detailed explanation of this technique can be found in, for example,][]{jones10a}.

To date, seven PNe with binary central stars have been studied in sufficient detail to reliably compare the inclinations of both binary orbital plane and nebular symmetry axis, with each and every system showing results consistent with the shaping influence of the central binary (see table \ref{tab:incl} for a list of these systems).

\begin{table}
\caption{The seven PNe with central binaries which have been the subject of detailed modelling for both the nebular morphology/inclination and binary.}
\label{tab:incl}
\begin{center}
{\small
\begin{tabular}{ccc}
\tableline
\noalign{\smallskip}
PN & Morphokinematical modelling & Binary modelling\\
\noalign{\smallskip}
\tableline
\noalign{\smallskip}
A~41 & \citet{jones10b} & \citet{bruch01}\\
A~63 & \citet{mitchell07} & \citet{bell94}\\
A~65 & \citet{huckvale13} & \cite{shimansky09}\\
HaTr~4 & \citet{tyndall12} & \cite{bodman12}\\
NGC~6337 & \citet{garcia-diaz09} & \citet{hillwig10}\\
NGC~6778 & \citet{guerrero12} & \citet{miszalski11b}\\
&& (Not modelled but eclipsing)\\
Sp~1 & \citet{jones12} & \citet{bodman12}\\
\noalign{\smallskip}
\tableline
\end{tabular}
}
\end{center}
\end{table}

\subsection{Pre-CE polar outflows?}
\label{sec:polar}

In addition to linking orientation of nebula and binary plane, morphokinematic modelling has revealed an interesting trend amongst those PNe with jet-like outflows - in the few cases studied, the jets do not appear coeval with the main nebula (see table \ref{tab:jets}).  In the majority of cases (though still small numbers), the jets appear older by roughly 1000 years.  While this could possibly be accounted for by interaction with the interstellar medium slowing the jets and making them appear older, it is unlikely to be the case for all systems.  As such, it is more likely that, while the main nebular shell is the product of the ejection of the CE, the jets arise from a period of mass transfer \emph{prior} to the formation of the CE.  Indeed, \citet{miszalski13b} found direct evidence of accretion onto the secondary in the central star system of The Necklace, probably originating from such a phase of pre-CE mass transfer, further supporting this hypothesis.

The idea that jets are formed before the onset of the CE is particularly interesting as it offers the prospect of being able to study the state of the binary just before the CE phase, while the current state is representative of that just after the CE (as the presence of the nebula indicates that the binary will not have had sufficient time to evolve substantially following the ejection).  This means that these systems offer a fantastic test-case for models of the poorly understood CE phase of close binary evolution.

\begin{table}[!ht]
\caption{The kinematical ages of PNe, with binary central stars, and their observed polar outflows.}
\label{tab:jets}
\begin{center}
{\small
\begin{tabular}{cccc}
\tableline
\noalign{\smallskip}
PN & PN age & Polar outflow age & Reference\\
& (yrs kpc$^{-1}$) & (yrs kpc$^{-1}$)\\
\noalign{\smallskip}
\tableline
\noalign{\smallskip}
Abell 63 & $3500\pm200$ & $5200\pm1200$ & \citet{mitchell07}\\
ETHOS~1 & $900\pm100$ & $1750\pm250$ & \citet{miszalski11a}\\
The Necklace & $1100\pm100$ & $2350\pm450$ & \citet{corradi11} \\
Fg~1 & $\sim2000$ & 2500--7000 & \citet{lopez93,boffin12} \\
NGC~6337 & $\sim9200$ & $\sim1000$ & \citet{garcia-diaz09} \\
\noalign{\smallskip}
\tableline
\end{tabular}
}
\end{center}
\end{table}

\begin{figure}
\includegraphics[width=\textwidth]{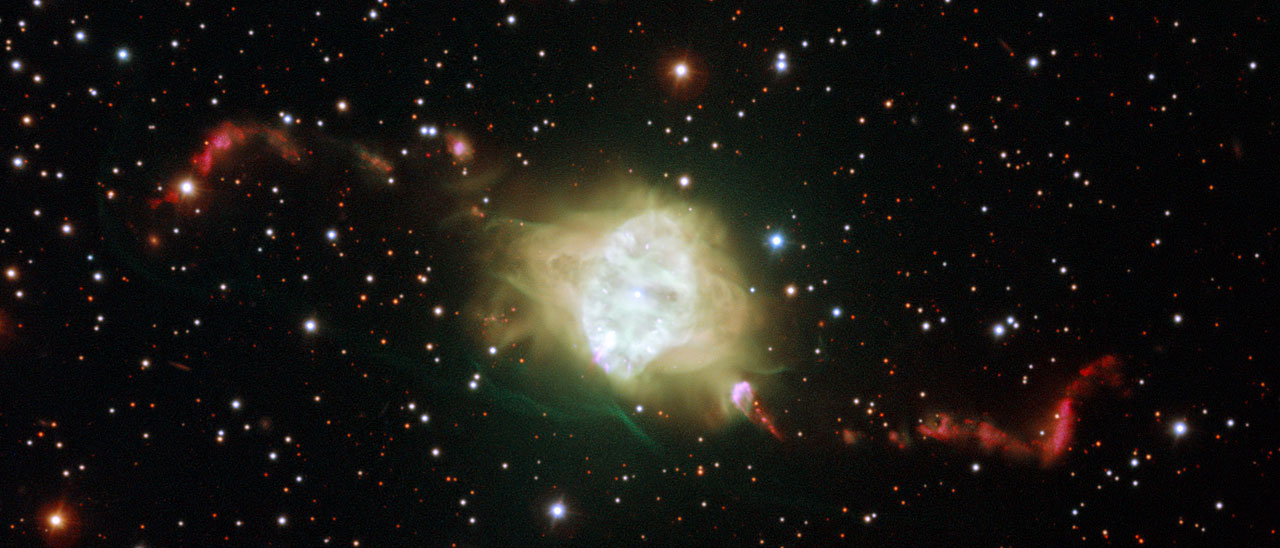}
\caption{The PN Fg~1 shown by \citet{boffin12} to host a binary central star, which was responsible for launching the spectacular episodic jets before to entering the CE phase (Image credit: ESO/H. Boffin).}
\label{fig:fg1}
\end{figure}

A great example of the power of PN to study the CE phase comes in the form of Fg~1 (Fig.\ \ref{fig:fg1}), a PN which has long been hypothesised to have been formed through binary interaction given its spectacular system of precessing jets \citep{boffin12}.  Based on the precession period of the jets \citep[determined by morphokinematic modelling;][]{lopez93} and hydrodynamic models of their formation \citep{raga09}, it was believed that the binary period should be of order 100--1000 years.  But, radial velocity monitoring revealed the central star to be a double degenerate system of period 1.19 days.  This, along with the difference in ages between jets and nebula, indicates that the jets were formed during a phase of mass transfer before the CE when the binary had a longer period.  As such, it is possible to somewhat reconstruct the history of this fascinating object.

The central star system of Fg~1 must have begun life as an unequal mass pair where the nebular progenitor was the lower mass component.  As the (now) secondary evolved off the main sequence (MS), the pair probably went through a symbiotic star phase, reducing the orbital period while avoiding a CE.  Ultimately, the secondary will have evolved to the white dwarf (WD) phase while the primary was still on the MS.  As the primary then evolved off the MS, it transferred mass to its companion forming the precessing jets in the process, before eventually entering into a CE \citep[the ejection of which formed the now visible PN;][]{boffin12}.  Interestingly, given the masses of the two WDs currently in the system, the progenitors would have had a pre-CE orbital separation (period$\sim$100--1000 years) greater than the radius of the progenitor whilst on the AGB, meaning that the pre-CE system must have had non-zero orbital eccentricity in order to enter the CE phase.  Hence, in this incredible system we are able to infer information about the evolution of the binary through the CE phase, a critically important and poorly understood process.

One of the nebulae listed in table \ref{tab:jets}, however, does not conform to this trend of jets older than nebular core -- NGC~6337 \citep{garcia-diaz09}.  In this object, the jets are younger than the nebula, indicating that the mass transfer/accretion episode that let to their formation occurred \emph{after} the CE was ejected.  While some theories of the CE do allow for continued accretion after its ejection, the jets are perhaps more likely the result of post-CE infall or even accretion from the secondary back onto the primary.  The latter is an intriguing possibility as many of the WD-MS post-CE systems seen in PNe are found to show inflated secondaries (inflated with respect to their expected MS radii), possibly a consequence of the accretion episode prior to the formation of (or even during) the CE phase.  It might therefore be possible that the secondary of NGC~6337 left the CE highly inflated, and shortly after proceeded to transfer matter back onto the primary leading to the formation of the jets.  Whatever the true origin of the jets, the system surely warrants further investigation.

By assuming that the jets are formed by an accretion disk threaded by a magnetic field and that the same magnetic field is responsible for launching and collimation of the jet, \citep{tocknell14} were able to indirectly measure the pre-CE (and post-CE, in the case of NGC~6778) magnetic fields in a selection of these objects for which ionised masses have been measured.  Thus showing that these objects are an excellent probe of CE evolution, allowing us to constrain many of the systemic parameters both before and after the CE phase.

\section{Long/intermediate-period binaries}

Thus far, discussion has been reserved for short period, post-CE binaries only, but these are not the only binary systems known to reside at the centre of PNe.  Another population of longer period binaries has been the subject of further study during recent years, the Barium star planetary nebulae \citep[LoTr~5, WeBo~1, A~70, Hen~2-39:][]{thevenin97,bond03,miszalski12,miszalski13}.  These PNe are known to contain an intermediate period binary of period $\sim$100--1000 days due to the detection of a cool, chemically polluted component to the central star spectrum in addition to the hot component (the nebular progenitor).  These are clearly formed by a binary interaction as the nebular progenitor ascends the Asymptotic Giant Branch donating matter to its less evolved companion, chemically enriching its surface in carbon or  s-process elements such as Barium \citep{boffin88}.  While these systems do not have measured periods or inclinations, their nebular morphologies indicate that all share a similar evolution, with all known systems displaying pronounced equatorial rings/waists and, where data of sufficient depth exists, faint lobes \citep{tyndall13}.

\section{Discussion}
Recent studies have clearly shown that binary central stars are responsible for the shaping of their host PNe, will all systems studied showing the predicted alignment between nebular symmetry axis and binary orbital plane.  Further detailed study of key objects has shown the possibility that different nebular components may trace different phases in the evolution of the central binary, with polar ejections occurring before the binary entered the CE phase.  In one particularly special case, it has been possible to trace the evolution of the central binary through two phases of interaction, with the first avoiding a CE while the second did not, resulting in the formation of a double degenerate binary core and a bipolar nebula with spectacular precessing jets.

In spite of this clear progress, any relationship between binary parameters and morphokinematical nebular parameters remains elusive, clearly requiring detailed study of a much broader sample of nebulae and central binaries.

\acknowledgements 
This work was co-funded under the Marie Curie Actions of the European Commission (FP7-COFUND). This research has made use of NASA's Astrophysics Data System Bibliographic Services. This research has made use of the SIMBAD database, operated at CDS, Strasbourg, France.

\bibliography{djones}

\end{document}